\documentclass[prc,aps,amsfonts,amssymb,floats,twocolumn]{revtex4}
\usepackage{graphics}
\usepackage{epsfig}

\begin{document}
\title{Probing the Low-Energy Electronic Structure of Complex Systems by ARPES}
\author{Andrea Damascelli}
\address{Department of Physics {\rm {\&}} Astronomy, University of
British Columbia, 6224 Agricultural Road, Vancouver, British
Columbia V6T\,1Z1, Canada}

\begin{abstract}
Angle-resolved photoemission spectroscopy (ARPES) is one of the
most direct methods of studying the electronic structure of
solids. By measuring the kinetic energy and angular distribution
of the electrons photoemitted from a sample illuminated with
sufficiently high-energy radiation, one can gain information on
both the energy and momentum of the electrons propagating inside a
material. This is of vital importance in elucidating the
connection between electronic, magnetic, and chemical structure of
solids, in particular for those complex systems which cannot be
appropriately described within the independent-particle picture.
The last decade witnessed significant progress in this technique
and its applications, thus ushering in a new era in photoelectron
spectroscopy; today, ARPES experiments with 2 meV energy
resolution and $0.2^\circ$ angular resolution are a reality even
for photoemission on solids. In this paper we will review the
fundamentals of the technique and present some illustrative
experimental results; we will show how ARPES can probe the
momentum-dependent electronic structure of solids providing
detailed information on band dispersion and Fermi surface, as well
as on the strength and nature of those many-body correlations
which may profoundly affect the one-electron excitation spectrum
and, in turn, determine the macroscopic physical properties.
\end{abstract}

\maketitle


\section{Introduction}
\label{sec:introduc}

Photoelectron spectroscopy is a general term that refers to all
those techniques based on the application of the photoelectric
effect originally observed by Hertz \cite{Hertz:1} and later
explained as a manifestation of the quantum nature of light by
Einstein \cite{EinsteinA:1}, who recognized that when light is
incident on a sample an electron can absorb a photon and escape
from the material with a maximum kinetic energy
$E_{kin}\!=\!h\nu\!-\!\phi$ (where $\nu$ is the photon frequency
and $\phi$, the material work function, is a measure of the
potential barrier at the surface that prevents the valence
electrons from escaping, and is typically 4-5 eV in metals). In
the following, we will show how the photoelectric effect also
provides us with deep insights into the quantum description of the
solid state. In particular, we will give a general overview of
angle-resolved photoemission spectroscopy (so-called ARPES), a
highly advanced spectroscopic method that allows the direct
experimental study of the momentum-dependent electronic band
structure of solids. For a further discussion of ARPES and other
spectroscopic techniques based on the detection of photoemitted
electrons, we refer the reader to the extensive literature
available on the subject
\cite{bachrach:1,BraunJ:thea-r,brundle:1,brundle:2,cardona:1,carlsonta:1,CourthsR:Phoec.,DamascelliA:Motios,Damascelli:RMP,EastmanDE:Phosm.,FeuerbacherB:Phoesc,feuerbacher:1,Grioni:1,HimpselFJ:Ang-re,hufner:1,InglesfieldJE:Eles.,kevan:1,Leckey:1,ley:1,lindau:1,lynch:1,mahan:1,margaritondo:1,nemoshkalenko,plummer:1,ShenZX:Elesps,SmithNV:Phopm.,SmithNV:Phos.,SmithKE:elesss,wendin:1,wertheim:1,WilliamsRH:Phosst}.

\begin{figure}[b!]
\centerline{\epsfig{figure=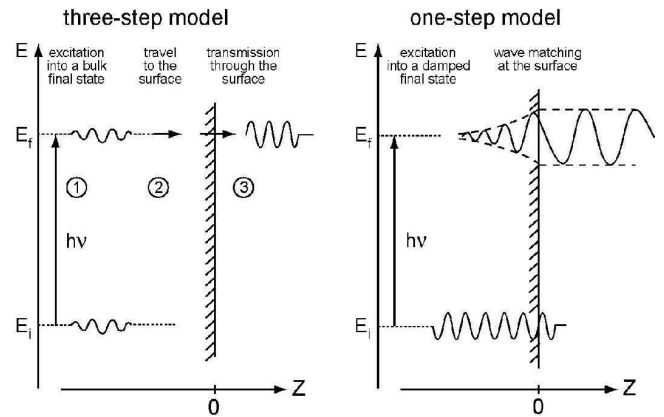,width=1\linewidth,clip=}}
\vspace{0cm} \caption{Pictorial representation of three-step and
one-step model description of the photoemission process (from
Ref.\,\onlinecite{hufner:1}). }\label{1vs3step}\end{figure}

\begin{figure}[t!]
\centerline{\epsfig{figure=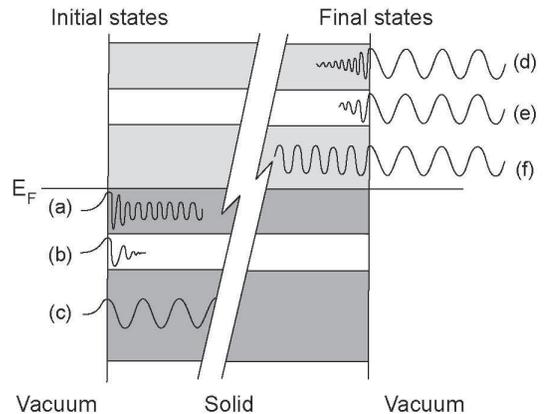,width=7cm,clip=}}
\vspace{0cm} \caption{Initial (left) and final (right) eigenstates
for the semi-infinite crystal. Left: (a) surface resonance; (b)
surface Shockley state situated in a gap of the bulk band
structure; (c) bulk Bloch state. Right: (d) surface resonance; (e)
in gap evanescent state; (f) bulk Bloch final state (from
Ref.\,\onlinecite{meinders:1}).}\label{eigenfunctions}\end{figure}

As we will see in detail throughout the paper and in particular in
Sec.\,\ref{sec:sudden}, due to the complexity of the photoemission
process in solids the quantitative analysis of the experimental
data is often performed under the assumption of the {\it
independent-particle picture} and of the {\it sudden
approximation} (i.e., disregarding the many-body interactions as
well as the relaxation of the system during the photoemission
itself). The problem is further simplified within the so-called
{\it three-step model} (Fig.\,\ref{1vs3step}a), in which the
photoemission event is decomposed in three independent steps:
optical excitation between the initial and final {\it bulk} Bloch
eigenstates, {\it travel} of the excited electron to the surface,
and escape of the photoelectron into vacuum after transmission
through the {\it surface} potential barrier. This is the most
common approach, in particular when photoemission spectroscopy is
used as a tool to map the electronic band structure of solids.
However, from the quantum-mechanical point of view photoemission
should not be described in terms of several independent events but
rather as a {\it one-step} process (Fig.\,\ref{1vs3step}b): in
terms of an optical transition (with probability given by
Eq.\,\ref{eq:wfi}) between initial and final states consisting of
many-body wave functions that obey appropriate boundary conditions
at the surface of the solid. In particular (see
Fig.\,\ref{eigenfunctions}), the initial state should be one of
the possible $N$-electron eigenstates of the semi-infinite
crystal, and the final state must be one of the eigenstates of the
ionized $(N\!-\!1)$-electron semi-infinite crystal; the latter has
also to include a component consisting of a propagating plane-wave
in vacuum (to account for the escaping photoelectron) with a
finite amplitude inside the crystal (to provide some overlap with
the initial state). Furthermore, as expressed by Eq.\,\ref{eq:wfi}
which does represent a complete one-step description of the
problem, in order for an electron to be photoemitted in vacuum not
only there must be a finite overlap between the amplitude of
initial and final states, but also the following energy and
momentum conservation laws for the impinging photon and the
$N$-electron system as a whole must be obeyed:
\begin{eqnarray}
E_{f}^N-E_{i}^N&=&h\nu \label{eq:totenegy}
\\
{\bf k}_{f}^N-{\bf k}_{i}^N&=&{\bf k}_{h\nu}
\label{eq:totmomentum}
\end{eqnarray}
\noindent
where the indexes $i$ and $f$ refer to initial and final state,
respectively, and ${\bf k}_{h\nu}$ is the momentum of the incoming
photon. Note that, in the following, in proceeding with the more
detailed analysis of the photoemission process as well as its
application to the study of the momentum-dependent electronic
structure of solids (in terms of both conventional band mapping as
well as many-body effects), we will mainly restrict ourselves to
the context of the three-step model and the sudden approximation.

\section{Kinematics of photoemission}
\label{sec:kine}

\begin{figure}[t!]
\centerline{\epsfig{figure=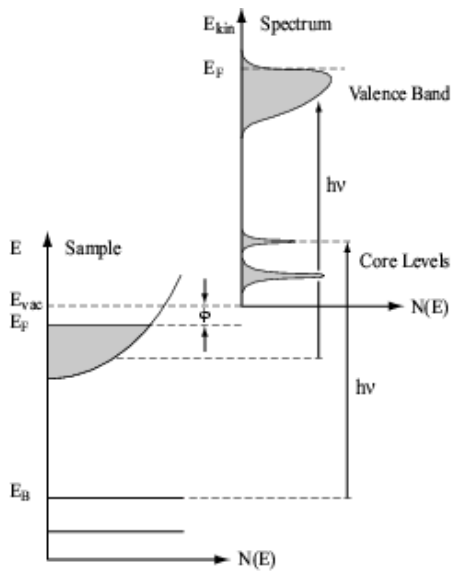,width=6.0cm,clip=}}
\vspace{0cm} \caption{Energetics of the photoemission process
(from Ref.\,\onlinecite{hufner:1}). The electron energy
distribution produced by the incoming photons, and measured as a
function of the kinetic energy $E_{kin}$ of the photoelectrons
(right), is more conveniently expressed in terms of the binding
energy $E_B$ (left) when one refers to the density of states in
the solid ($E_{0,B}\!=\!0$ at
$E_F$).}\label{energetics}\end{figure}

\begin{figure}[b!]
\centerline{\epsfig{figure=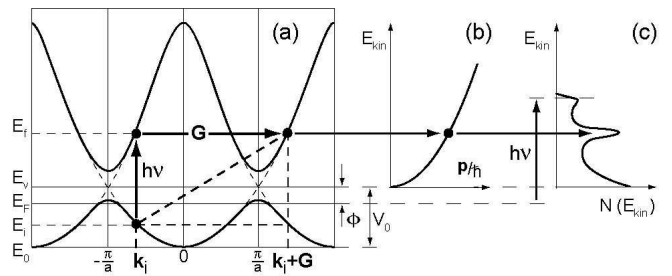,width=1\linewidth,clip=}}
\vspace{0cm} \caption{Kinematics of the photoemission process
within the three-step nearly-free-electron final state model: (a)
direct optical transition in the solid (the lattice supplies the
required momentum); (b) free-electron final state in vacuum; (c)
corresponding photoelectron spectrum, with a background due to the
scattered electrons ($E_{0,i,f}\!=\!0$ at $E_F$). From
Ref.\,\onlinecite{pillo:1}.}\label{kinematics}\end{figure}

The energetics and kinematics of the photoemission process are
shown in Fig.\ref{energetics} and\,\ref{kinematics}, while the
geometry of an ARPES experiment is sketched in
Fig.\ref{geometry}a. A beam of monochromatized radiation supplied
either by a gas-discharge lamp or by a synchrotron beamline is
incident on a sample (which has to be a properly aligned single
crystal in order to perform {\it angle} or, equivalently, {\it
momentum}-resolved measurements). As a result, electrons are
emitted by photoelectric effect and escape in vacuum in all
directions. By collecting the photoelectrons with an electron
energy analyzer characterized by a finite acceptance angle, one
measures their kinetic energy $E_{kin}$ for a given emission
direction. This way, the wave vector or momentum ${\bf K}\!=\!{\bf
p}/\hbar$ of the photoelectrons {\it in vacuum} is also completely
determined: its modulus is given by $
K\!=\!\sqrt{2mE_{kin}}/\hbar$ and its components parallel (${\bf
K_{||}}\!=\!{\bf K}_{x}+{\bf K}_{y}$) and perpendicular (${\bf
K_{\perp}}\!=\!{\bf K}_{z}$) to the sample surface are obtained in
terms of the polar ($\vartheta$) and azimuthal ($\varphi$)
emission angles defined by the experiment:
\begin{eqnarray}
K_x&=&\frac{1}{\hbar}\sqrt{2mE_{kin}}\sin\vartheta\cos\varphi
\\
K_y&=&\frac{1}{\hbar}\sqrt{2mE_{kin}}\sin\vartheta\sin\varphi
\\
K_z&=&\frac{1}{\hbar}\sqrt{2mE_{kin}}\cos\vartheta\
\label{eq:Kvac}
\end{eqnarray}
\noindent
The goal is then to deduce the electronic dispersion relations
$E({\bf k})$ for the solid left behind, i.e. the relation between
binding energy $E_B$ and momentum ${\bf k}$ for the electrons
propagating {\it inside} the solid, starting from $E_{kin}$ and
${\bf K}$ measured for the photoelectrons {\it in vacuum}.
In order to do that, one has to exploit the total energy and
momentum conservation laws (Eq.\,\ref{eq:totenegy}
and\,\ref{eq:totmomentum}, respectively).

Within the non-interacting electron picture, it is particularly
straightforward to take advantage of the energy conservation law
and relate, as pictorially described in Fig.\,\ref{energetics},
the kinetic energy of the photoelectron to the binding energy
$E_B$ of the electronic-state inside the solid:
\begin{equation}
E_{kin}=h\nu-\phi-|E_B| \label{eq:enegy}
\end{equation}
More complex, as we will discuss below, is to gain full knowledge
of the crystal electronic momentum ${\bf k}$. Note, however, that
the photon momentum can be neglected in Eq.\,\ref{eq:totmomentum}
at the low photon energies most often used in ARPES experiments
($h\nu\!<\!100$\,eV), as it is much smaller than the typical
Brillouin-zone dimension $2\pi/a$ of a solid (see
Sec.\,\ref{sec:stateart} for more details). Thus, as shown in
Fig.\,\ref{kinematics} within the three-step model description
(see also Sec.\,\ref{sec:sudden}), the optical transition between
the bulk initial and final states can be described by a vertical
transition in the {\it reduced-zone scheme} (${\bf k}_{f}\!-\!{\bf
k}_{i}\!=\!0$), or equivalently by a transition between
momentum-space points connected by a reciprocal-lattice vector
{\bf G} in the {\it extended-zone scheme} (${\bf k}_{f}\!-\!{\bf
k}_{i}\!=\!{\bf G}$). In regard to Eq.\,\ref{eq:totenegy}
and\,\ref{eq:totmomentum} and the deeper meaning of the
reciprocal-lattice vector {\bf G} note that, as emphasized by
Mahan in his seminal paper on the theory of photoemission in
simple metals \cite{MahanGD:Thepsm}, ``{\it in a
nearly-free-electron gas, optical absorption may be viewed as a
two-step process. The absorption of the photon provides the
electron with the additional energy it needs to get to the excited
state. The crystal potential imparts to the electron the
additional momentum it needs to reach the excited state. This
momentum comes in multiples of the reciprocal-lattice vectors
{\boldmath$G$}. So in a reduced zone picture, the transitions are
vertical in wave-vector space. But in photoemission, it is more
useful to think in an extended-zone scheme.}'' On the contrary in
an infinite crystal with no periodic potential (i.e., a truly
free-electron gas scenario lacking of any periodic momentum
structure), no ${\bf k}$-conserving transition is possible in the
limit ${\bf k}_{h\nu}\!=\!0$, as one cannot go from an initial to
a final state along the same unperturbed free-electron parabola
without an external source of momentum. In other words, direct
transitions are prevented because of the lack of appropriate final
states (as opposed to the periodic case of
Fig.\,\ref{kinematics}).
Then again the problem would be quite different if the surface was
more realistically taken into account, as in a one-step model
description of a semi-infinite crystal. In fact, while the surface
does not perturb the translational symmetry in the $x$-$y$ plane
and ${\bf k_{\|}}$ is conserved to within a reciprocal lattice
vector ${\bf G_{\|}}$, due to the abrupt potential change along
the $z$ axis the perpendicular momentum ${\bf k}_{\bot}$ is not
conserved across the sample surface (i.e., ${\bf k_{\perp}}$ is
not a good quantum number except than deeply into the solid,
contrary to ${\bf k_{||}}$). Thus, the surface can play a direct
role in momentum conservation, delivering the necessary momentum
for indirect transitions even in absence of the crystal potential
(i.e., the so-called {\it surface photoelectric effect}; see also
Eq.\,\ref{eq:wfi} and the related discussion).

Reverting to the three-step model {\it direct-transition}
description of Fig.\,\ref{kinematics}, the transmission through
the sample surface is obtained by matching the bulk Bloch
eigenstates inside the sample to free-electron plane waves in
vacuum. Because of the translational symmetry in the $x$-$y$ plane
across the surface, from these matching conditions it follows that
the parallel component of the electron momentum is actually
conserved in the process:
\begin{equation}
{\bf k}_{\|}={\bf
K}_{\|}=\frac{1}{\hbar}\sqrt{2mE_{kin}}\cdot\sin\vartheta
\label{eq:momentum}
\end{equation}
\noindent
where ${\bf k}_{\|}$ is the component parallel to the surface of
the electron crystal momentum in the {\it extended}-zone scheme
(upon going to larger $\vartheta$ angles, one actually probes
electrons with ${\bf k}_{\|}$ lying in higher-order Brillouin
zones; by subtracting the corresponding reciprocal-lattice vector
${\bf G_{\|}}$, the {\it reduced} electron crystal momentum in the
first Brillouin zone is obtained). As for the determination of
${\bf k}_{\bot}$, which is not conserved but is also needed in
order to map the electronic dispersion $E({\bf k})$ vs the total
crystal wave vector ${\bf k}$, a different approach is required.
As a matter of fact, several specific experimental methods for
absolute three dimensional band mapping have been developed
\cite{hufner:1,StrocovVN:Newmab,StrocovVN:Absbmb}, which however
are rather complex and require additional and/or complementary
experimental data. Alternatively, the value of ${\bf k}_{\bot}$
can be determined if some {\it a priori} assumption is made for
the dispersion of the electron final states involved in the
photoemission process; in particular, one can either use the
results of band structure calculations, or adopt a
nearly-free-electron description for the final bulk Bloch states:
\begin{equation}
E_f({\bf k})= \frac{\hbar^2{\bf k}^2}{2m}-|E_0|=
\frac{\hbar^2({\bf k_\|}^2+{\bf k_\perp}^2)}{2m}-|E_0|
\label{eq:free}
\end{equation}
\noindent
where once again the electron momenta are defined in the
extended-zone scheme, and $E_0$ corresponds to the bottom of the
valence band as indicated in Fig.\,\ref{kinematics} (note that
both $E_0$ and $E_f$ are referenced to the Fermi energy $E_F$,
while $E_{kin}$ is referenced to the vacuum level $E_{v}$).
Because $E_{f}\!=\!E_{kin}\!+\!\phi$ and $\hbar^2{\bf
k}_{\|}^2/2m\!=\!E_{kin}\sin^2\vartheta$, which follow from
Fig.\,\ref{kinematics} and  Eq.\,\ref{eq:momentum}, one obtains
from Eq.\,\ref{eq:free}:
\begin{equation}
{\bf k}_{\perp}=
\frac{1}{\hbar}\sqrt{2m(E_{kin}\cos^2\vartheta+V_0)}
\label{eq:kperp}
\end{equation}
\noindent
Here $V_0\!=\!|E_0|\!+\!\phi$ is the {\it inner potential}, which
corresponds to the energy of the bottom of the valence band
referenced to vacuum level $E_{v}$. From Eq.\,\ref{eq:kperp} and
the measured values of $E_{kin}$ and $\vartheta$, if $V_0$ is also
known, one can then obtain the corresponding value of ${\bf
k}_{\perp}$. As for the determination of $V_0$, three methods are
generally used: (i) optimize the agreement between theoretical and
experimental band mapping for the occupied electronic state; (ii)
set $V_0$ equal to the theoretical zero of the muffin tin
potential used in band structure calculations; (iii) infer $V_0$
from the experimentally observed periodicity of the dispersion
$E({\bf k}_{\perp})$. The latter is actually the most convenient
method as the experiment can be realized by simply detecting the
photoelectrons emitted along the surface normal (i.e., ${\bf
K}_{\|}\!=\!0$) while varying the incident photon energy and, in
turn, the energy $E_{kin}$ of the photoelectrons and thus ${\bf
K}_{z}$ (see Eq.\,\ref{eq:Kvac}).
\begin{figure}[t!]
\centerline{\epsfig{figure=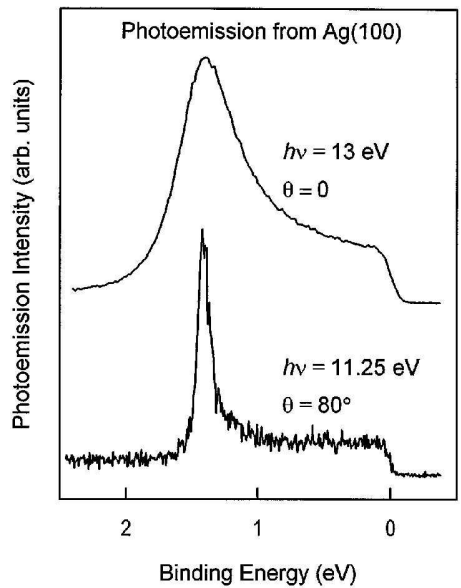,width=6cm,clip=}}
\vspace{0cm} \caption{Normal and grazing emission ARPES spectra
from Ag(100) measured with photon energies specifically chosen to
give rise to peaks with the same binding  energy (from
Ref.\,\onlinecite{HansenED:Observ}).}\label{silver}\end{figure}
Note that the nearly-free electron approximation for the final
states is expected to work well for materials in which the Fermi
surface has a simple spherical (free-electron-like) topology such
as in the alkali metals, and for high-energy final states in which
case the crystal potential is a small perturbation (eventually the
final-state bands become so closely spaced in energy to form a
continuum, and the details of the final states become
unimportant). However this approximation is often used also for
more complicated systems, even if the initial states are not free
electron-like.

A particular case in which the uncertainty in ${\bf k}_{\bot}$ is
less relevant is that of the low-dimensional systems characterized
by an anisotropic electronic structure and, in particular, a
negligible dispersion along the $z$ axis (i.e., the surface
normal, see Fig.\,\ref{geometry}a). The electronic dispersion is
then almost exclusively determined by ${\bf k}_{\|}$ (as in the
case of many transition metal oxides, such as for example the
two-dimensional copper oxide superconductors
\cite{Damascelli:RMP}). As a result, one can map out in detail the
electronic dispersion relations $E({\bf k})$ simply by tracking,
as a function of ${\bf K}_{\|}$, the energy position of the peaks
detected in the ARPES spectra for different take-off angles (as in
Fig.\,\ref{geometry}b, where both direct and inverse photoemission
spectra for a single band dispersing through the Fermi energy
$E_F$ are shown). Furthermore, as an additional bonus of the lack
of $z$ dispersion, one can directly identify the width of the
photoemission peaks with the lifetime of the photohole
\cite{SMITHNV:PHOLQL}, which contains information on the intrinsic
correlation effects of the system and is formally described by the
imaginary part of the electron self energy (see
Sec.\,\ref{sec:spectral}). On the contrary, in 3D systems the
linewidth contains contributions from both photohole and
photoelectron lifetimes, with the latter reflecting final state
scattering processes and thus the finite probing depth; as a
consequence, isolating the intrinsic many-body effects becomes a
much more complicated problem.

What just discussed for the lifetime can be easily seen from the
expression for the FWHM of an ARPES lineshape for a single
nearly-free electron-like band \cite{SMITHNV:PHOLQL}:
\begin{equation}
\Gamma=\frac
{\frac{\Gamma_i}{|v_{i\perp}|}+\frac{\Gamma_f}{|v_{f\perp}|}}
{\left|\frac{1}{v_{i\perp}}\left[1-\frac{mv_{i\|}\sin
^2\vartheta}{\hbar k_{\|}}\right] -
\frac{1}{v_{f\perp}}\left[1-\frac{mv_{f\|}\sin ^2\vartheta}{\hbar
k_{\|}}\right]\right|} \label{eq:gammatot}
\end{equation}
\noindent
Here $\Gamma_f$ and $\Gamma_i$ are the inverse lifetime of
photoelectron and photohole in the final and initial states,
respectively, and $v_i$ and $v_f$ are the corresponding group
velocities (e.g., $\hbar v_{i\perp}\!=\!\partial E_i/\partial
k_{\perp}$). Note in particular that: (i) for initial states very
close to $E_F$, $\Gamma_i\!\longrightarrow\!0$ and the linewidth
reflects only the lifetime of the final state $\Gamma_f$; (ii)
Eq.\,\ref{eq:gammatot} simplifies considerably in the case of a
material characterized by a two dimensional electronic structure,
for which $|v_{i\perp}|\!\simeq\!0$; as a result, the final-state
lifetime contribution vanishes:
\begin{equation}
\Gamma=\frac{\Gamma_i} {\left|1-\frac{mv_{i\|}\sin
^2\vartheta}{\hbar k_{\|}}\right|}\equiv C\, \Gamma_i
\label{eq:gamma}
\end{equation}
\noindent
Furthermore, depending on the sign of $v_{i\|}$, the measured
linewidth can be compressed or expanded with respect to the
intrinsic value of the inverse lifetime $\Gamma_i$. The two
limiting cases mentioned above are beautifully exemplified by the
data from the three-dimensional system Ag(100) presented in
Fig.\,\ref{silver} \cite{HansenED:Observ}. While the normal
incidence spectrum is dominated by $\Gamma_f\!\gg\!\Gamma_i$ and
is extremely broad, the grazing incidence data from a momentum
space region characterized by $v_{i\perp}\!=\!0$, $v_{i\|}\!<\!0$
and large, and $k_{\|}$ small (which result in a compression
factor $C\!=\!0.5$), exhibit a linewidth which is even narrower
than the intrinsic inverse lifetime $\Gamma_i$. Note that this
does not imply any fundamental violation of the basic principles
of quantum mechanics, but is just a direct consequence of the
kinematics constrains of the photoemission process.

\section{Three-step model and sudden approximation}
\label{sec:sudden}

To develop a formal description of the photoemission process, one
has to calculate the transition probability $w_{fi}$ for an
optical excitation between the $N$-electron ground state
$\Psi_i^N$ and one of the possible final states $\Psi_f^N$. This
can be approximated by Fermi's golden rule:
\begin{equation}
w_{fi}=\frac{2\pi}{\hbar}|\langle\Psi_f^N|H_{int}|\Psi_i^N\rangle|^2\delta(E_f^N-E_i^N-h\nu)
\label{eq:wfi}
\end{equation}
where $E_i^N\!=\!E_i^{N-1}\!-\!E_B^{\bf k}$ and
$E_f^N\!=\!E_f^{N-1}\!+\!E_{kin}$ are the initial and final-state
energies of the $N$-particle system ($E_B^{\bf k}$ is the binding
energy of the photoelectron with kinetic energy $E_{kin}$ and
momentum ${\bf k}$). The interaction with the photon is treated as
a perturbation given by:
\begin{equation}
H_{int}=\frac{e}{2mc}({\bf A}\!\cdot\!{\bf p}+{\bf p}\!\cdot\!{\bf
A})= \frac{e}{mc}\,{\bf A}\!\cdot\!{\bf p} \label{eq:hint}
\end{equation}
where {\bf p} is the electronic momentum operator and {\bf A} is
the electromagnetic vector potential (note that the gauge
$\Phi\!=\!0$ was chosen for the scalar potential $\Phi$, and the
quadratic term in ${\bf A}$ was dropped because in the linear
optical regime it is typically negligible with respect to the
linear terms). In Eq.\,\ref{eq:hint} we also made use of the
commutator relation $[{\bf p},{\bf
A}]\!=\!-i\hbar\nabla\!\cdot\!{\bf A}$ and dipole approximation
[i.e., {\bf A} constant over atomic dimensions and therefore
$\nabla\!\cdot\!{\bf A}\!=\!0$, which holds in the ultraviolet].
Although this is a routinely used approximation, it should be
noted that $\nabla\!\cdot\!{\bf A}$ might become important at the
{\it surface} where the electromagnetic fields may have a strong
spatial dependence. This surface photoemission contribution, which
is proportional to ($\varepsilon-1$) where $\varepsilon$ is the
medium dielectric function, can interfere with the bulk
contribution resulting in asymmetric lineshapes for the bulk
direct-transition peaks
\cite{feuerbacher:1,MillerT:Intbbs,HansenED:SurpA,HansenED:Ovetsp}.
At this point, a more rigorous approach is to proceed with the
so-called {\it one-step model} (Fig.\,\ref{1vs3step}b), in which
photon absorption, electron removal, and electron detection are
treated as a single coherent process
\cite{mitchell:1,makinson:1,buckingham:1,MahanGD:Thepsm,SchaichWL:Modctt,FeibelmanPJ:Phos-C,PendryJB:Thep.,pendry:1,LiebschA:Thepla,liebsch:1,LindroosM:Sursa-,LindroosM:novdmF,BansilA:Ang-re,BansilA:Mateet,BansilA:Impmet}.
In this case bulk, surface, and vacuum have to be included in the
Hamiltonian describing the crystal, which implies that not only
bulk states have to be considered but also surface and evanescent
states, and surface resonances (see Fig.\,\ref{eigenfunctions}).
Note that, under the assumption $\nabla\!\cdot\!{\bf A}\!=\!0$,
from Eq.\,\ref{eq:hint} and the commutation relation $[H_0,{\bf
p}]\!=\!i\hbar \nabla V$ (where $H_0\!=\!{\bf p}^2/2m\!+\!V$ is
the unperturbed Hamiltonian of the semi-infinite crystal) it
follows that the matrix elements appearing in Eq.\,\ref{eq:wfi}
are proportional to $\langle\Psi_f^N|{\bf A}\!\cdot\!\nabla
V|\Psi_i^N\rangle$. This shows explicitly that for a true
free-electron like system it would be impossible to satisfy
simultaneously energy and momentum conservation laws inside the
material because there $\nabla V\!=\!0$. The only region where
electrons could be photoexcited is at the surface where $\partial
V/\partial z\!\neq\!0$, which gives rise to the so-called {\it
surface photoelectric effect}. However, due to the complexity of
the one-step model, photoemission data are usually discussed
within the three-step model (Fig.\,\ref{1vs3step}a), which
although purely phenomenological has proven to be rather
successful \cite{FanHY:1,berglund:1,FeibelmanPJ:Phos-C}. Within
this approach, the photoemission process is subdivided into three
independent and sequential steps:
\begin{figure*}[t!]
\centerline{\epsfig{figure=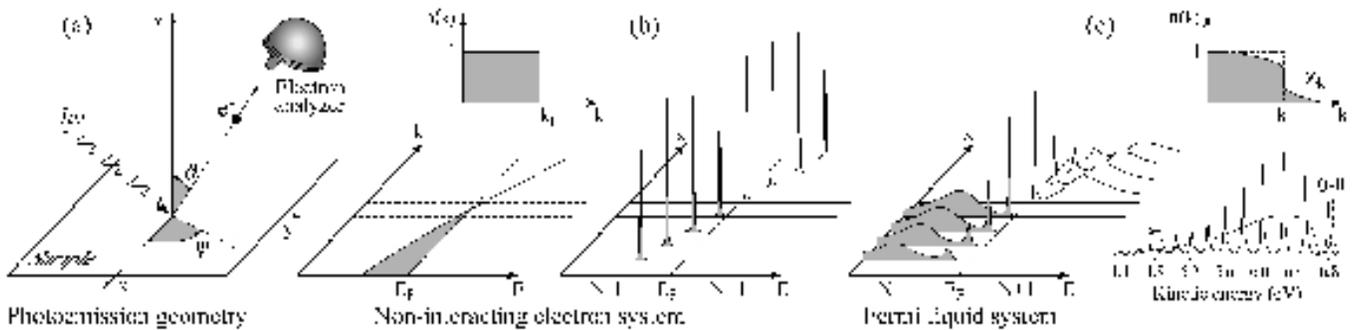,width=1\linewidth,clip=}}
\caption{(a) Geometry of an ARPES experiment; the emission
direction of the photoelectron is specified by the polar
($\vartheta$) and azimuthal ($\varphi$) angles. Momentum resolved
one-electron removal and addition spectra for: (b) a
non-interacting electron system (with a single energy band
dispersing across the Fermi level); (c) an interacting Fermi
liquid system. The corresponding ground-state ($T\!=\!0$ K)
momentum distribution function $n({\bf k})$ is also shown. (c)
Bottom right: photoelectron spectrum of gaseous hydrogen and the
ARPES spectrum of solid hydrogen developed from the gaseous one
(from Ref.\,\onlinecite{sawatzky:1} and\,\onlinecite{meinders:1}).
}\label{geometry}\end{figure*}
\begin{itemize}
\item[(i)]{Optical excitation of the electron in the {\it bulk}.}
\item[(ii)]{Travel of the excited electron to the surface.}
\item[(iii)]{Escape of the photoelectron into vacuum.}
\end{itemize}
 The total photoemission intensity is then given by the product
of three independent terms: the total probability for the optical
transition, the scattering probability for the travelling
electrons, and the transmission probability through the surface
potential barrier. Step (i) contains all the information about the
intrinsic electronic structure of the material and will be
discussed in detail below. Step (ii) can be described in terms of
an effective mean free path, proportional to the probability that
the excited electron will reach the surface without scattering
(i.e, with no change in energy and momentum). The inelastic
scattering processes, which determine the surface sensitivity of
photoemission (see Sec.\,\ref{sec:stateart}), give rise to a
continuous background in the spectra which is usually ignored or
subtracted. Step (iii) is described by a transmission probability
through the surface, which depends on the energy of the excited
electron and the material work function $\phi$ (in order to have
any finite escape probability the condition $\hbar^2 {\bf
k}_{\perp}^2/2m\!\geq\!|E_0|\!+\!\phi$ must be satisfied).

In evaluating step (i), and therefore the photoemission intensity
in terms of the transition probability $w_{fi}$, it would be
convenient to factorize the wavefunctions in Eq.\,\ref{eq:wfi}
into photoelectron and $(N\!-\!1)$-electron terms, as we have done
for the corresponding energies. This however is far from trivial
because during the photoemission process itself the system will
relax. The problem simplifies within the {\it sudden
approximation}, which is extensively used in many-body
calculations of the photoemission spectra from interacting
electron systems, and is in principle applicable only to high
kinetic-energy electrons. In this limit, the photoemission process
is assumed to be {\it sudden}, with no post-collisional
interaction between the photoelectron and the system left behind
(in other words, an electron is instantaneously removed and the
effective potential of the system changes discontinuously at that
instant). The final state $\Psi^{N}_{f}$ can then be written as:
\begin{equation}
\Psi_f^N={\mathcal{A}}\, \phi_f^{\bf k}\, \Psi^{N-1}_f
\end{equation}
where $\mathcal{A}$ is an antisymmetric operator that properly
antisymmetrizes the $N$-electron wavefunction so that the Pauli
principle is satisfied, $\phi_f^{\bf k}$ is the wavefunction of
the photoelectron with momentum ${\bf k}$, and $\Psi^{N-1}_{f}$ is
the final state wavefunction of the $(N\!-\!1)$-electron system
left behind, which can be chosen as an excited state with
eigenfunction $\Psi^{N-1}_{m}$ and energy $E^{N-1}_m$. The {\it
total} transition probability is then given by the sum over {\it
all} possible excited states $m$. Note, however, that the sudden
approximation is inappropriate for low kinetic energy
photoelectrons, which may need longer than the system response
time to escape into vacuum. In this case, the so-called {\it
adiabatic limit}, one can no longer factorize $\Psi_f^N$ in two
independent parts and the detailed screening of photoelectron and
photohole has to be taken into account \cite{GadzukJW:Excedc}.

For the initial state, let us first assume for simplicity that
$\Psi^{N}_{i}$ is a single Slater determinant (i.e., Hartree-Fock
formalism), so that we can write it as the product of a
one-electron orbital $\phi_i^{\bf k}$ and an $(N\!-\!1)$-particle
term:
\begin{equation}
\Psi_i^N={\mathcal{A}}\, \phi_i^{\bf k}\, \Psi^{N-1}_i
\end{equation}
More generally, however, $\Psi^{N-1}_{i}$ should be expressed as
$\Psi^{N-1}_i\!=\!c_{\bf k}\Psi^{N}_i$, where $c_{\bf k}$ is the
annihilation operator for an electron with momentum $\bf k$. This
also shows that $\Psi^{N-1}_{i}$ is {\it not} an eigenstate of the
$(N\!-\!1)$ particle Hamiltonian, but is just what remains of the
$N$-particle wavefunction after having pulled out one electron. At
this point, we can write the matrix elements in Eq.\,\ref{eq:wfi}
as:
\begin{equation}
\langle\Psi_f^N|H_{int}|\Psi_i^N\rangle\!=\!\langle\phi_f^{\bf
k}|H_{int}|\phi_i^{\bf k}\rangle
\langle\Psi^{N-1}_{m}|\Psi^{N-1}_{i}\rangle
\end{equation}
where $\langle\phi_f^{\bf k}|H_{int}|\phi_i^{\bf
k}\rangle\!\equiv\!M_{f,i}^{\bf k}$ is the one-electron dipole
matrix element, and the second term is the $(N\!-\!1)$-electron
overlap integral. Here, we replaced $\Psi_f^{N-1}$ with an
eigenstate $\Psi_m^{N-1}$, as discussed above. The total
photoemission intensity measured as a function of $E_{kin}$ at a
momentum ${\bf k}$, namely $I({\bf
k},E_{kin})\!=\!\sum_{f,i}w_{f,i}$, is then proportional to:
\begin{equation}
\sum_{f,i}|M_{f,i}^{\bf k}|^2 \sum_{m}
|c_{m,i}|^2\delta(E_{kin}\!+\!E_{m}^{N-1}\!-E_i^N\!-\!h\nu)
\label{eq:photocurrent}
\end{equation}
where
$|c_{m,i}|^2\!=\!|\langle\Psi^{N-1}_{m}|\Psi^{N-1}_{i}\rangle|^2$
is the probability that the removal of an electron from state $i$
will leave the $(N\!-\!1)$-particle system in the excited state
$m$. From here we see that, if
$\Psi^{N-1}_{i}\!=\!\Psi^{N-1}_{m_0}$ for one particular
$m\!=\!m_0$, the corresponding $|c_{m_0,i}|^2$ will be unity and
all the others $c_{m,i}$ zero; in this case, if also $M_{f,i}^{\bf
k}\!\neq\!0$, the ARPES spectra will be given by a delta function
at the Hartree-Fock orbital energy $E_B^{\bf k}\!=\!-\epsilon_{\bf
k}$, as shown in Fig.\,\ref{geometry}b (i.e., non-interacting
particle picture). In the strongly correlated systems, however,
many of the $|c_{m,i}|^2$ will be different from zero because the
removal of the photoelectron results in a strong change of the
system effective potential and, in turn, $\Psi^{N-1}_{i}$ will
have an overlap with many of the eigenstates $\Psi^{N-1}_{m}$.
Therefore, the ARPES spectra will not consist of single delta
functions but will show a main line and several satellites
according to the number of excited states $m$ created in the
process (Fig.\,\ref{geometry}c).

What discussed above is very similar to the situation encountered
in photoemission from molecular hydrogen \cite{siegbahn:1} in
which not simply a single peak but many lines separated by few
tenths of eV from each other are observed (solid line in
Fig.\,\ref{geometry}c, bottom right). These so-called `shake-up'
peaks correspond to the excitations of the different vibrational
states of the H$_2^+$ molecule. In the case of solid hydrogen
(dashed line in Fig.\,\ref{geometry}c, bottom right), as discussed
by \cite{sawatzky:1}, the vibrational excitations would develop in
a broad continuum while a sharp peak would be observed for the
fundamental transition (from the ground state of the H$_2$ to the
one of the H$_2^+$ molecule). Note that the fundamental line would
also be the only one detected in the adiabatic limit, in which
case the $(N\!-\!1)$-particle system is left in its ground state.

\section{One-particle spectral function} \label{sec:spectral}

In the discussion of photoemission on solids, and in particular on
the correlated electron systems in which many $|c_{m,i}|^2$ in
Eq.\,\ref{eq:photocurrent} are different from zero, the most
powerful and commonly used approach is based on the Green's
function formalism
\cite{Abrikosov:1,hedin:1,FetterAL:1,mahan:2,economou:1,rickayzen:1}.
 In this context, the propagation of a single electron in a many-body
system is described by the {\it time-ordered} one-electron Green's
function ${\mathcal{G}}(t-t^\prime)$, which can be interpreted as
the probability amplitude that an electron added to the system in
a Bloch state with momentum ${\bf k}$ at a time zero will still be
in the same state after a time $|t\!-\!t^\prime|$. By taking the
Fourier transform, ${\mathcal{G}}(t\!-\!t^\prime)$ can be
expressed in energy-momentum representation resulting in
${\mathcal{G}}({\bf k},\omega)\!=\!G^+({\bf
k},\omega)\!+\!G^-({\bf k},\omega)$, where $G^+({\bf k},\omega)$
and $G^-({\bf k},\omega)$ are the one-electron addition and
removal Green's function, respectively. At $T\!=\!0$:
\begin{equation}
G^\pm({\bf
k},\omega)=\sum_{m}\frac{|\langle\Psi^{N\pm1}_m|c^\pm_{\bf k}
|\Psi^N_i\rangle|^2}{\omega- E_m^{N\pm 1}+ E^N_i\pm i\eta}
\end{equation}
where the operator $c_{\bf k}^+\!=\!c_{{\bf k}\sigma}^\dag$
($c_{\bf k}^-\!=\!c_{{\bf k}\sigma}^{ }$) creates (annihilates) an
electron with energy $\omega$, momentum ${\bf k}$, and spin
$\sigma$ in the $N$-particle initial state $\Psi^N_i$, the
summation runs over all possible $({N\!\pm\!1})$-particle
eigenstates $\Psi_m^{N\pm1}$ with eigenvalues $E_m^{N\pm1}$, and
$\eta$ is a positive infinitesimal (note also that from here on we
will take $\hbar\!=\!1$). In the limit $\eta\!\rightarrow\!0^+$
one can make use of the identity
$(x\!\pm\!i\eta)^{-1}\!=\!{\mathcal{P}}(1/x)\!\mp\!i\pi\delta(x)$,
where $\mathcal{P}$ denotes the principle value, to obtain the
{\it one-particle spectral function}  $A({\bf
k},\omega)\!=\!A^+({\bf k},\omega)\!+\!A^-({\bf
k},\omega)\!=\!-(1/\pi){\rm Im}\,G({\bf k},\omega)$, with:
\begin{equation}
A^\pm({\bf
k},\omega)\!=\!\!\!\sum_{m}|\langle\Psi^{N\pm1}_m\!|c^\pm_{\bf k}
|\Psi^N_i\rangle|^2 \delta(\omega\!-\!E_m^{N\pm 1}\!\!+\!E^N_i)
\label{eq:akw0}
\end{equation}
and $G({\bf k},\omega)\!=\!G^+({\bf k},\omega)\!+\![G^{-}({\bf
k},\omega)]^*$, which defines the {\it retarded} Green's function.
Note that $A^-({\bf k},\omega)$ and $A^+({\bf k},\omega)$ define
the one-electron removal and addition spectra which one can probe
with direct and inverse photoemission, respectively. This is
evidenced, for the direct case, by the comparison between the
expression for $A^-({\bf k},\omega)$ and
Eq.\,\ref{eq:photocurrent} for the photoemission intensity (note
that in the latter $\Psi^{N-1}_i\!=\!c_{\bf k}\Psi^{N}_i$ and the
energetics of the photoemission process has been explicitly
accounted for). Finite temperatures effect can be taken into
account by extending the Green's function formalism just
introduced to $T\!\neq\!0$ (see, e.g.,
Ref.\,\onlinecite{mahan:2}). In the latter case, by invoking once
again the sudden approximation the intensity measured in an ARPES
experiment on a 2D  single-band system can be conveniently written
as:
\begin{equation}
I({\bf k},\omega)=I_0({\bf k},\nu,{\bf A})f(\omega)A({\bf
k},\omega) \label{eq:ikw} \label{eq:intone}
\end{equation}
where ${\bf k}\!=\!{\bf k}_\parallel$ is the in-plane electron
momentum, $\omega$ is the electron energy with respect to the
Fermi level, and $I_0({\bf k},\nu,{\bf A})$ is proportional to the
squared one-electron matrix element $|M_{f,i}^{\bf k}|^2$ and
therefore depends on the electron momentum, and on the energy and
polarization of the incoming photon. We also introduced the Fermi
function $f(\omega)\!=\!(e^{\omega/k_BT}\!+\!1)^{-1}$ which
accounts for the fact that direct photoemission probes only the
occupied electronic states. Note that in Eq.\,\ref{eq:ikw} we
neglected the presence of any extrinsic background and the
broadening due to the energy and momentum resolution, which
however have to be carefully considered when performing a
quantitative analysis of the ARPES spectra (see
Sec.\,\ref{sec:matele} and Eq.\,\ref{eq:ikwreal}).

The corrections to the Green's function originating from
electron-electron correlations can be conveniently expressed in
terms of the electron {\it proper self energy} $\Sigma({\bf
k},\omega)\!=\!\Sigma^{\prime}({\bf
k},\omega)\!+\!i\Sigma^{\prime\prime}({\bf k},\omega)$. Its real
and imaginary part contain all the information on the energy
renormalization and lifetime, respectively, of an electron with
band energy $\epsilon_k$ and momentum ${\bf k}$ propagating in a
many-body system. The Green's and spectral functions expressed in
terms of the self energy are then given by:
\begin{equation}
G({\bf k},\omega)=\frac{1}{\omega-\epsilon_{\bf k}-\Sigma({\bf
k},\omega)} \label{eq:gwk}
\end{equation}
\vspace{-.55cm}
\begin{equation} A({\bf
k},\omega)=-\frac{1}{\pi}\frac{\Sigma^{\prime\prime}({\bf
k},\omega)}{[\omega-\epsilon_{\bf k}-\Sigma^{\prime}({\bf
k},\omega)]^2+[\Sigma^{\prime\prime}({\bf k},\omega)]^2}
\label{eq:akw1}
\end{equation}
Because $G(t,t^\prime)$ is a linear response function to an
external perturbation, the real and imaginary parts of its Fourier
transform $G({\bf k},\omega)$ have to satisfy causality and,
therefore, are related by Kramers-Kronig relations. This implies
that if the full $A({\bf k},\omega)\!=\!-(1/\pi){\rm Im}\,G({\bf
k},\omega)$ is available from photoemission and inverse
photoemission, one can calculate ${\rm Re}\,G({\bf k},\omega)$ and
then obtain both the real and imaginary parts of the self energy
directly from Eq.\,\ref{eq:gwk}. However, due to the lack of
high-quality inverse photoemission data, this analysis is usually
performed only using ARPES spectra by taking advantage of certain
approximations (such as, e.g., particle-hole symmetry within a
narrow energy range about $E_F$ \cite{NormanMR:Exttes}).

In general, the exact calculation of $\Sigma({\bf k},\omega)$ and,
in turn, of $A({\bf k},\omega)$ is an extremely difficult task. In
the following, as an example we will briefly consider the
interacting FL case \cite{landau:1,landau:2,landau:3}. Let us
start from the trivial $\Sigma({\bf k},\omega)\!=\!0$
non-interacting case.  The $N$-particle eigenfunction $\Psi^N$ is
a single Slater determinant and we always end up in a single
eigenstate when removing or adding an electron with momentum ${\bf
k}$. Therefore, $G({\bf
k},\omega)\!=\!1/(\omega\!-\!\epsilon_k\!\pm\!i\eta)$ has only one
pole for each ${\bf k}$, and $A({\bf
k},\omega)\!=\!\delta(\omega\!-\!\epsilon_k)$ consists of a single
line at the band energy $\epsilon_k$ (as shown in
Fig.\,\ref{geometry}b). In this case, the occupation numbers
$n_{{\bf k}\sigma}^{ }\!=\!c_{{\bf k}\sigma}^\dag c_{{\bf
k}\sigma}^{ }$ are good quantum numbers and for a metallic system
the {\it momentum distribution} [i.e., the expectation value
$n({\bf k})\!\equiv\!\langle n_{{\bf k}\sigma}\rangle$, quite
generally independent of the spin $\sigma$ for nonmagnetic
systems], is characterized by a sudden drop from 1 to 0 at ${\bf
k}\!=\!{\bf k}_F$ (Fig.\,\ref{geometry}b, top), which defines a
sharp Fermi surface. If we now switch on the electron-electron
correlation adiabatically, (so that the system remains at
equilibrium), any particle added into a Bloch state has a certain
probability of being scattered out of it by a collision with
another electron, leaving the system in an excited state in which
additional electron-hole pairs have been created. The momentum
distribution $n({\bf k})$ will now show a discontinuity smaller
than 1 at ${\bf k}_F$ and a finite occupation probability for
${\bf k}\!>\!{\bf k}_F$ even at $T\!=\!0$ (Fig.\,\ref{geometry}c,
top). As long as $n({\bf k})$ shows a finite discontinuity $Z_{\bf
k}\!>\!0$ at ${\bf k}\!=\!{\bf k}_F$, we can describe the
correlated Fermi sea in terms of well defined {\it
quasiparticles}, i.e. electrons {\it dressed} with a manifold of
excited states, which are characterized by a pole structure
similar to the one of the non-interacting system but with
renormalized energy $\varepsilon_{\bf k}$ and mass $m^*$, and a
finite lifetime $\tau_{\bf k}\!=\!1/\Gamma_{\bf k}$. In other
words, the properties of a FL are similar to those of a free
electron gas with damped quasiparticles. As the bare-electron
character of the quasiparticle or pole strength (also called
coherence factor) is $Z_{\bf k}\!<\!1$ and the total spectral
weight must be conserved (see Eq.\,\ref{eq:intakw}), we can
separate $G({\bf k},\omega)$ and $A({\bf k},\omega)$ into a {\it
coherent} pole part and an {\it incoherent} smooth part without
poles \cite{pines:1}:
\begin{equation}
G({\bf k},\omega)=\frac{Z_{\bf k}}{\omega-\varepsilon_{\bf
k}+i\Gamma_{\bf k}}+G_{inch} \label{eq:gfl}
\end{equation}
\vspace{-.55cm}
\begin{equation} A({\bf
k},\omega)=Z_{\bf k}\frac{\Gamma_{\bf k}/\pi}
{(\omega-\varepsilon_{\bf k})^2+\Gamma_{\bf k}^2}+A_{inch}
\end{equation}
where $Z_{\bf
k}\!=\!(1\!-\!\frac{\partial\Sigma^{\prime}}{\partial\omega})^{-1}$,
$\varepsilon_{\bf k}\!=\!Z_{\bf k}(\epsilon_{\bf
k}\!+\!\Sigma^{\prime})$, $\Gamma_{\bf k}\!=\!Z_{\bf
k}|\Sigma^{\prime\prime}|$, and the self energy and its
derivatives are evaluated at $\omega\!=\!\varepsilon_{\bf k}$. It
should be emphasized that the FL description is valid only in
proximity to the Fermi surface and rests on the condition
$\varepsilon_{\bf k}\!-\!\mu\!\gg\!|\Sigma^{\prime\prime}|$ for
small $(\omega\!-\!\mu)$ and $({\bf k}\!-\!{\bf k}_F)$.
Furthermore, $\Gamma_{\bf k}\!\propto\![(\pi
k_BT)^2\!+\!(\varepsilon_{\bf k}\!-\!\mu)^2]$ for a FL system in
two or more dimensions \cite{luttinger:2,pines:1}, although
additional logarithmic corrections should be included in the
two-dimensional case \cite{HodgesC:EffFsg}. By comparing the
electron removal and addition spectra for a FL of quasiparticles
with those of a non-interacting electron system (in the lattice
periodic potential), the effect of the self-energy correction
becomes evident (see Fig.\,\ref{geometry}c and \,\ref{geometry}b,
respectively). The quasiparticle peak has now a finite lifetime
(due to $\Sigma^{\prime\prime}$), and it sharpens up rapidly thus
emerging from the broad incoherent component upon approaching the
Fermi level, where the lifetime is infinite corresponding to a
well defined quasiparticle [note that the coherent and incoherent
part of $A({\bf k},\omega)$ represent the main line and satellite
structure discussed in the previous section and shown in
Fig.\,\ref{geometry}c, bottom right].
\begin{figure}[t!]
\centerline{\epsfig{figure=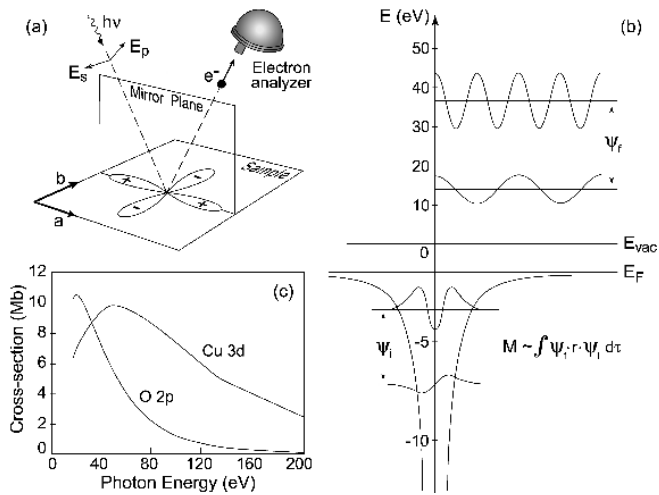,width=1\linewidth,clip=}}
\caption{(a) Mirror plane emission from a $d_{x^2-y^2}$ orbital.
(b) Sketch of the optical transition between atomic orbitals with
different angular momenta (the harmonic oscillator wavefunctions
are here used for simplicity) and free electron wavefunctions with
different kinetic energies (from Ref.\,\onlinecite{hufner:1}). (c)
Calculated photon energy dependence of the photoionization
cross-sections for Cu 3$d$ and O 2$p$ atomic levels (from
Ref.\,\onlinecite{YehJJ:Atospc}).} \label{matrelem}
\end{figure}
Furthermore, the peak position is shifted with respect to the bare
band energy $\epsilon_{\bf k}$ (due to $\Sigma^{\prime}$): as the
quasiparticle mass is larger than the band mass because of the
dressing ($m^*\!>\!m$), the total dispersion (or bandwidth) will
be smaller ($|\varepsilon_{\bf k}|\!<\!|\epsilon_{\bf k}|$).

Among the general properties of the spectral function there are
also several sum rules. A fundamental one, which in discussing the
FL model was implicitly used to state that $\int\!d\omega
A_{ch}\!=\!Z_{\bf k}$ and $\int\!d\omega A_{inch}\!=\!1\!-\!Z_{\bf
k}$ (where $A_{ch}$ and $A_{inch}$ refer to coherent and
incoherent parts of the spectral function, respectively), is the
following:
\begin{equation}
\int_{-\infty}^{+\infty}d\omega A({\bf k},\omega)=1
\label{eq:intakw}
\end{equation}
which reminds us that $A({\bf k},\omega)$ describes the
probability of removing/adding an electron with  momentum {\bf k}
and energy $\omega$ to a many-body system. However, as it also
requires the knowledge of the electron addition part of the
spectral function, it is not so useful in the analysis of ARPES
data. A sum rule more relevant to this task is:
\begin{equation}
\int_{-\infty}^{+\infty}d\omega f(\omega)A({\bf k},\omega)=n({\bf
k}) \label{eq:nk}
\end{equation}
which solely relates the one-electron removal spectrum to the
momentum distribution $n({\bf k})$. When electronic correlations
are important and the occupation numbers are no longer good
quantum numbers, the discontinuity at ${\bf k}_F$ is reduced (as
discussed for the FL case) but a drop in $n({\bf k})$ is usually
still observable even for strong correlations \cite{nozieres:1}.
By tracking in $k$-space the {\it loci} of steepest descent of the
experimentally determined $n({\bf k})$, i.e. maxima in
$|\nabla\!_{\bf k}\,n({\bf k})|$, one may thus identify the Fermi
surface even in those correlated systems exhibiting particularly
complex ARPES features. However, great care is necessary in making
use of Eq.\,\ref{eq:nk} because the integral of Eq.\,\ref{eq:ikw}
does not give just $n({\bf k})$ but rather $I_0({\bf k},\nu,{\bf
A})n({\bf k})$ \cite{Damascelli:RMP}.

\section{Matrix elements and finite resolution effects}
\label{sec:matele}

As discussed in the previous section and summarized by
Eq.\,\ref{eq:ikw}, ARPES directly probes the one-particle spectral
function $A({\bf k},\omega)$. However, in extracting quantitative
information from the experiment, not only the effect of the matrix
element term $I_0({\bf k},\nu,{\bf A})$ has to be taken into
account, but also the finite experimental resolution and the
extrinsic continuous background due to the secondaries (those
electrons which escape from the solid after having suffered
inelastic scattering events and, therefore, with a reduced
$E_{kin}$). The latter two effects may be explicitly accounted for
by considering a more realistic expression for the photocurrent
$I({\bf k},\omega)$:
\begin{equation}
\int\!\!d\tilde{\omega}d\tilde{{\bf k}}\,I_0(\tilde{{\bf
k}},\!\nu,\!{\bf A})f(\tilde{\omega})A(\tilde{{\bf
k}},\!\tilde{\omega})R(\omega\!-\!\tilde{\omega})Q({\bf
k}\!-\!\tilde{{\bf k}})\!+\!B \label{eq:ikwreal}
\end{equation}
which consists of the convolution of Eq.\,\ref{eq:ikw} with energy
($R$) and momentum ($Q$) resolution functions [$R$ is typically a
Gaussian, $Q$ may be more complicated], and of the background
correction B. Of the several possible forms for the background
function B \cite{hufner:1}, two are more frequently used: (i) the
step-edge background (with three parameters for height, energy
position, and width of the step-edge), which reproduces the
background observed all the way to $E_F$ in an unoccupied region
of momentum space;
\begin{figure}[b!]
\centerline{\epsfig{figure=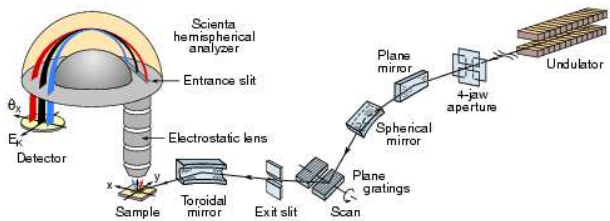,width=1\linewidth,clip=}}
\caption{(Color). Beamline equipped with a plane grating
monochromator and a 2D position-sensitive electron analyzer.}
\label{scienta}
\end{figure}
(ii) the Shirley background
$B_{Sh}(\omega)\!\propto\!\int^\mu_\omega d\omega^\prime
P(\omega^\prime)$, which allows to extract from the measured
photocurrent $I(\omega)\!=\!P(\omega)\!+\!c_{Sh} B_{Sh}(\omega)$
the contribution $P(\omega)$ of the unscattered electrons (with
only the parameter $c_{Sh}$ \cite{ShirleyDA:Hig-re}).

Let us now very briefly illustrate the effect of the matrix
element term $I_0({\bf k},\nu,{\bf A})\!\propto\!|M_{f,i}^{\bf
k}|^2$, which is responsible for the dependence of the
photoemission data on photon energy and experimental geometry, and
may even result in complete suppression of the intensity
\cite{gobeligw:1,DietzE:Polda-,HermansonJ:Fin-st,EberhardtW:Dipsro}.
By using the commutation relation $\hbar{\bf p}/m\!=\!-i[{\bf
x},H]$, we can write $|M^{\bf k}_{f,i}|^2\!\propto\!|\langle
\phi_f^{\bf k}|\mbox{\boldmath$\varepsilon$}\!\cdot\!{\bf
x}|\phi_i^{\bf k}\rangle |^2$, where {\boldmath$\varepsilon$} is a
unit vector along the polarization direction of the vector
potential {\bf A}.  As in Fig.\,\ref{matrelem}a, let us consider
photoemission from a $d_{x^2-y^2}$ orbital, with the detector
located in the mirror plane (when the detector is out of the
mirror plane, the problem is more complicated because of the lack
of an overall well defined even/odd symmetry). In order to have
non vanishing photoemission intensity, the whole integrand in the
overlap integral must be an even function under reflection with
respect to the mirror plane. Because odd parity final states would
be zero everywhere on the mirror plane and therefore also at the
detector, the final state wavefunction $\phi_f^{\bf k}$ itself
must be even. In particular, at the detector the photoelectron is
described by an even parity plane-wave state $e^{i{\bf kr}}$ with
momentum in the mirror plane and fronts orthogonal to it
\cite{HermansonJ:Fin-st}. In turn, this implies that
$(\mbox{\boldmath$\varepsilon$}\!\cdot\!{\bf x})|\phi_i^{\bf
k}\rangle$ must be even. In the case depicted in
Fig.\,\ref{matrelem}a where $|\phi_i^{\bf k}\rangle$ is also even,
the photoemission process is symmetry allowed for {\bf A} even or
in-plane (i.e., $\mbox{\boldmath$\varepsilon$}_p\!\cdot\!{\bf x}$
depends only on in-plane coordinates and is therefore even under
reflection with respect to the plane) and forbidden for {\bf A}
odd or normal to the mirror plane (i.e.,
$\mbox{\boldmath$\varepsilon$}_s\!\cdot\!{\bf x}$ is odd as it
depends on normal-to-the-plane coordinates). For a generic initial
state of either even or odd symmetry with respect to the mirror
plane, the polarization conditions resulting in an overall even
matrix element can be summarized as:
\begin{equation}
\left\langle{\phi_f^{\bf k}}\right|{\bf A}\!\cdot\!{\bf
p}\left|{\phi_i^{\bf k}}\right\rangle \left\{{\matrix{{\phi_i^{\bf
k}\,\,even\,\,\left\langle + \right|+\left| + \right\rangle
\,\,\Rightarrow \,\,{\bf A}\,\,even} \vspace{0.1cm}\cr
{\!\!\!\phi_i^{\bf k}\,\,odd\,\,\,\,\,\left\langle +
\right|-\left| - \right\rangle \,\,\Rightarrow \,\,{\bf
A}\,\,odd}\cr }}\right.
\end{equation}

In order to discuss the photon energy dependence, from
Eq.\,\ref{eq:hint} and by considering a plane wave $e^{i{\bf kr}}$
for the photoelectron at the detector, one may more conveniently
write $|M^{\bf
k}_{f,i}|^2\!\propto\!|(\mbox{\boldmath$\varepsilon$}\!\cdot\!{\bf
k})\langle\phi_i^{\bf k}|e^{i{\bf kr}}\rangle |^2$.
\begin{figure}[t!]
\centerline{\epsfig{file=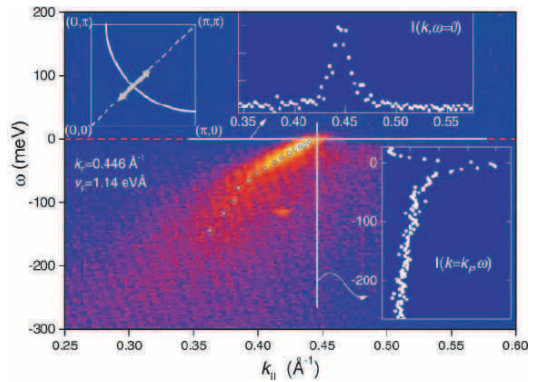,clip=,angle=0,width=0.8\linewidth}}
\caption{(Color). Energy ($\omega$) versus momentum (${\bf
k}_\parallel$) image plot of the photoemission intensity from
Bi$_2$Sr$_2$CaCu$_2$O$_{8+\delta}$ along (0,0)-($\pi$,$\pi$). This
$k$-space cut was taken across the Fermi surface (see sketch of
the 2D Brillouin zone) and allows a direct visualization of the
photohole spectral function $A({\bf k},\omega)$ (weighted by Fermi
distribution and matrix elements): the quasiparticle dispersion
can be clearly followed up to $E_F$, as emphasized by the white
circles. Energy scans at constant momentum (right) and momentum
scans at constant energy (top) define {\it energy distribution
curves} (EDCs) and {\it momentum distribution curves} (MDCs),
respectively (from Ref.\,\onlinecite{VallaT:Eviqcb}).}
\label{valla1}
\end{figure}
The overlap integral, as sketched in Fig.\,\ref{matrelem}b,
strongly depends on the details of the initial state wavefunction
(peak position of the radial part and oscillating character of
it), and on the wavelength of the outgoing plane wave. Upon
increasing the photon energy, both $E_{kin}$ and {\bf k} increase,
and $M^{\bf k}_{f,i}$ changes in a non-necessarily monotonic
fashion (see Fig.\,\ref{matrelem}c, for the Cu 3$d$ and the O 2$p$
atomic case). In fact, the photoionization cross section is
usually characterized by one minimum in free atoms, the so-called
Cooper minimum \cite{Cooperjw:1}, and a series of them in solids
\cite{MolodtsovSL:Coomtp}.

\section{State-of-the-art photoemission} \label{sec:stateart}

The configuration of a generic angle-resolved photoemission
beamline is shown in Fig.\,\ref{scienta}. A beam of white
radiation is produced in a wiggler or an undulator (these
so-called `insertion devices' are the straight sections of the
electron storage ring where radiation is produced); the light is
then monochromatized at the desired photon energy by a grating
monochromator, and is focused on the sample. Alternatively, a
gas-discharge lamp can be used as a radiation source (once
properly monochromatized, to avoid complications due to the
presence of different satellites and refocused to a small spot
size, essential for high angular resolution). However, synchrotron
radiation offers important advantages: it covers a wide spectral
range (from the visible to the X-ray region) with an intense and
highly polarized continuous spectrum, while a discharge lamp
provides only a few resonance lines at discrete energies.
Photoemitted electrons are then collected by the analyzer, where
kinetic energy and emission angle are determined (the whole system
is in ultra-high vacuum at pressures lower than
5$\times\!10^{-11}$ torr).

A conventional hemispherical analyzer consists of a multi-element
electrostatic input lens, a hemispherical deflector with entrance
and exit slits, and an electron detector (i.e., a channeltron or a
multi-channel detector). The heart of the analyzer is the
deflector which consists of two concentric hemispheres (of radius
$R_1$ and $R_2$). These are kept at a potential difference $\Delta
V$, so that only those electrons reaching the entrance slit with
kinetic energy within a narrow range centered at
$E_{pass}\!=\!e\Delta V/(R_1/R_2\!-\!R_2/R_1)$ will pass through
this hemispherical capacitor, thus reaching the exit slit and then
the detector. This way it is possible to measure the kinetic
energy of the photoelectrons with an energy resolution given by
$\Delta E_a\!=\!E_{pass}(w/R_0\!+\!\alpha^2\!/4)$, where
$R_0\!=\!(R_1\!+\!R_2)/2$, $w$ is the width of the entrance slit,
and $\alpha$ is the acceptance angle. The role of the
electrostatic lens is that of decelerating and focusing the
photoelectrons onto the entrance slit. By scanning the lens
retarding potential one can effectively record the photoemission
intensity versus the photoelectron kinetic energy. One of the
innovative characteristics of the state-of-the-art analyzer is the
two-dimensional position-sensitive detector consisting of two
micro-channel plates and a phosphor plate in series, followed by a
CCD camera. In this case, no exit slit is required: the electrons,
spread apart along the $Y$ axis of the detector
(Fig.\,\ref{scienta}) as a function of their kinetic energy due to
the travel through the hemispherical capacitor, are detected
simultaneously [in other words, a range of electron energies is
dispersed over one dimension of the detector and can be measured
in parallel; scanning the lens voltage is in principle no longer
necessary, at least for narrow energy windows (a few percent of
$E_{pass}$)]. Furthermore, contrary to a conventional electron
spectrometer in which the momentum information is averaged over
all the photoelectrons within the acceptance angle (typically
$\pm1^\circ$), state-of-the-art 2D position-sensitive electron
analyzers can be operated in angle-resolved mode, which provides
energy-momentum information not only at a single $k$-point but
along an extended cut in $k$-space.
\begin{figure}[t!]
\centerline{\epsfig{figure=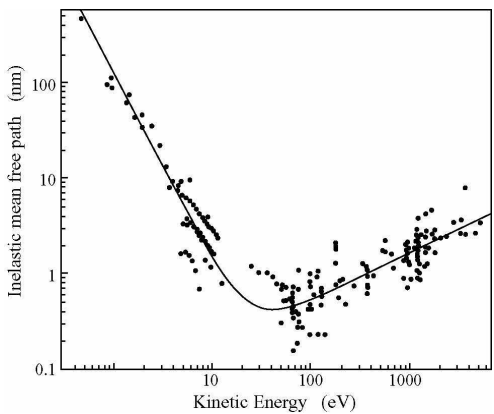,width=6.5cm,clip=}}
\caption{Kinetic energy dependence of the `universal' mean free
path for excited electrons in solids (from
Ref.\,\onlinecite{SeahMP:Quaess}).} \label{mfp}
\end{figure}
In particular, the photoelectrons within an angular window of
$\sim\!14^\circ$ along the direction defined by the analyzer
entrance slit are focused on different $X$ positions on the
detector (Fig.\,\ref{scienta}). It is thus possible to measure
multiple energy distribution curves simultaneously for different
photoelectron angles, obtaining a 2D snapshot of energy versus
momentum (Fig.\,\ref{valla1}).

State-of-the-art spectrometers typically allow for energy and
angular resolutions of approximately a few meV and $0.2^\circ$,
respectively. Taking as example the transition metal oxides and in
particular the cuprate superconductors (for which
$2\pi/a\!\simeq\!1.6$ \AA$^{-1}$), one can see from
Eq.\,\ref{eq:momentum} that 0.2$^\circ$ corresponds to $\sim$0.5\%
of the Brillouin zone size, for the 21.2 eV photons of the
HeI$\alpha$ line typically used in ARPES systems equipped with a
gas-discharge lamp.
\begin{figure}[b!]
\centerline{\epsfig{figure=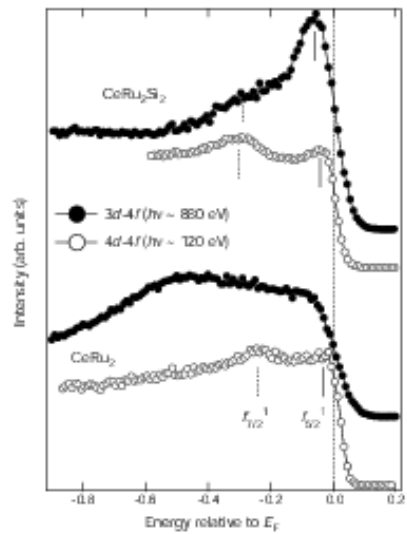,width=5.8cm,clip=}}
\vspace{0cm} \caption{High energy angle-integrated resonance
photoemission data from Ce compounds at $T\!=\!20$ K (from
Ref.\,\onlinecite{SekiyamaA:Probsc}).}\label{suga}\end{figure}
In the case of a beamline, to estimate the total energy resolution
one has to take into account also $\Delta E_{m}$ of the
monochromator, which can be adjusted with entrance and exit slits
(the ultimate resolution a monochromator can deliver is given by
its resolving power $R\!=\!E/\Delta E_{m}$; it can be as good as
1-2 meV for 20 eV photons but worsens upon increasing the photon
energy).
\begin{figure*}[t!]
\centerline{\epsfig{figure=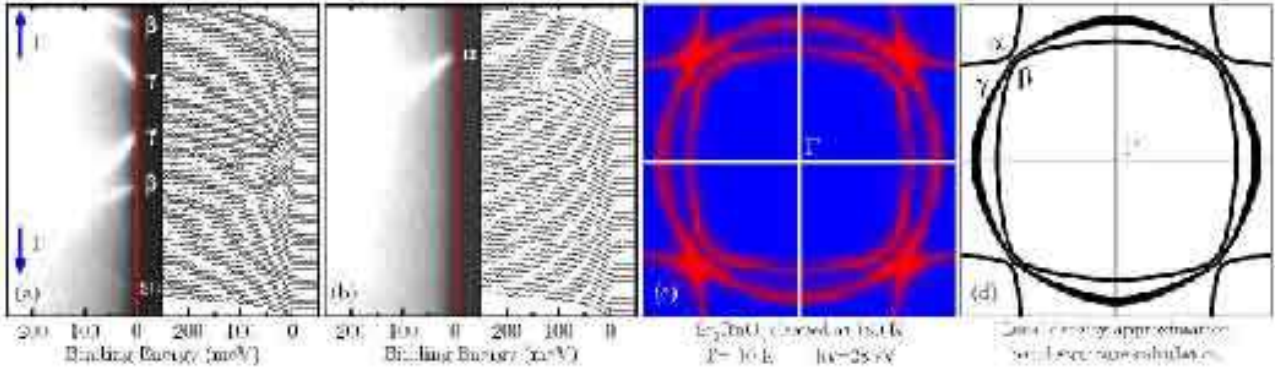,width=0.95\linewidth,clip=}}
\vspace{0cm} \caption{(Color). ARPES spectra and corresponding
intensity plot from Sr$_2$RuO$_4$ along (a) $\Gamma$-M, and (b)
M-X.  (c) Measured and (d) calculated \cite{MazinII:Fersfi} Fermi
surface. All data were taken at 10\,K on a Sr$_2$RuO$_4$ single
crystal cleaved at 180\,K (from Ref.\,
\onlinecite{DamascelliA:Fersss}).} \label{srofs}
\end{figure*}
To maximize the signal intensity at the desired total $\Delta E$,
monochromator and analyzer should be operated at comparable
resolutions. As for the momentum resolution $\Delta{\bf k}_{\|}$,
note that from Eq.\,\ref{eq:momentum} and neglecting the
contribution due to the finite energy resolution one can write:
\begin{equation}
\Delta{\bf k}_{\|}\simeq
\sqrt{2mE_{kin}/\hbar^2}\cdot\cos\vartheta\!\cdot\!\Delta\vartheta
\label{eq:res}
\end{equation}
where $\Delta\vartheta$ is the finite acceptance angle of the
electron analyzer. From Eq.\,\ref{eq:res} it is clear that the
momentum resolution is better at lower photon energy (i.e., lower
$E_{kin}$), and larger polar angles $\vartheta$ (one can
effectively improve the momentum resolution by extending the
measurements to momenta outside the first Brillouin zone).

Because at lower photon energies it is possible to achieve higher
energy and momentum resolution, most of the ARPES experiments are
performed in the ultraviolet (typically for $h\nu\!<\!100$ eV). An
additional advantage is that at low photon energies one can
disregard the photon momentum ${\bf k}_{h\nu}\!=\!2\pi/\lambda$ in
Eq.\,\ref{eq:momentum}, as for 100 eV photons the momentum is 0.05
\AA$^{-1}$ (only  3\% of the Brillouin zone size, by taking again
the cuprates as an example), and at 21.2 eV (HeI$\alpha$) it is
only 0.008 \AA$^{-1}$ (0.5\% of the zone). If on the contrary the
photon momentum is not negligible, the photoemission process does
not involve vertical transitions and $\kappa$ must be explicitly
taken into account in Eq.\,\ref{eq:momentum}. For example, for
1487 eV photons (the Al K$_{\alpha}$ line commonly used in X-ray
photoemission) ${\bf k}_{h\nu}\!\simeq\!0.76$\,\AA$^{-1}$, which
corresponds to 50\% of the zone size.

A major drawback of working at low photon energies is the extreme
surface sensitivity. As shown in Fig.\,\ref{mfp}, the mean free
path for unscattered photoelectrons is characterized by a minimum
of approximately 5 \AA\ at 20-100 eV kinetic energies
\cite{SeahMP:Quaess}, which are typical values in ARPES
experiments. This means that a considerable fraction of the total
photoemission intensity will be representative of the topmost
surface layer, especially on systems characterized by a large
structural/electronic anisotropy. Therefore, ARPES experiments
have to be performed on atomically clean and well-ordered systems,
which implies that atomically {\it fresh} and {\it flat} surfaces
have to be `prepared' immediately prior to the experiment in
ultra-high vacuum conditions (typically at pressures lower than
5$\times\!10^{-11}$ torr). Even then, however, because of the
lower atomic coordination at the surface, the coexistence of bulk
and surface electronic states, and the possible occurrence of
chemical and/or structural surface instabilities, photoemission
data may not always be representative of the intrinsic bulk
electronic structure. In order address with this issue, great care
has to be taken also over the structural and chemical
characterization of the sample surface, which can be done
independently by low-energy electron diffraction (LEED) and
core-level X-ray photoemission spectroscopy (XPS), respectively
(either prior to or during the ARPES experiments). In this regard
it has to be emphasized that, although the ultimate resolutions
are not as good as in the UV regime, the sensitivity to bulk over
surface electronic states can be enhanced (see Fig.\,\ref{mfp}) by
performing the ARPES experiments in the soft X-ray regime
(500-1500 eV). The significance of this approach is well
exemplified by recent angle-integrated resonance photoemission
experiments performed on Ce compounds \cite{SekiyamaA:Probsc}.
These Kondo systems are characterized by a very different degree
of hybridization between the 4$f$ electronic states and other
valence bands: the hybridization is stronger the larger the Kondo
temperature $T_K$. However, although CeRu$_2$Si$_2$ and CeRu$_2$
are characterized by very different $T_K$ (approximately 22 and
1000 K, respectively), earlier photoemission studies reported
similar spectra for the Ce 4$f$ electronic states. By performing
angle-integrated high resolution photoemission experiments at the
3$d$-4$f$ ($h\nu\!\simeq\!880$ eV, $\Delta E\!\simeq\!100$ meV)
and 4$d$-4$f$ ($h\nu\!\simeq\!120$ eV, $\Delta E\!\simeq\!50$ meV)
resonances (see Fig.\,\ref{suga}), it was observed that, while the
spectra for the two compounds are indeed qualitatively similar at
120 eV photon energy, they are remarkably different at 880 eV. As
the photoelectron mean free path increases from approximately 5 to
almost 20 \AA\ upon increasing the photon energy from 120 to 880
eV (Fig.\,\ref{mfp}), it was concluded that the 4$d$-4$f$ spectra
mainly reflect the surface 4$f$ electronic states. These are
different from those of the bulk and are not representative of the
intrinsic electronic properties of the two compounds, which are
more directly probed at 880 eV: the 3$d$-4$f$ spectra show a
prominent structure corresponding to the tail of a Kondo peak in
CeRu$_2$Si$_2$, and a broader feature reflecting the more
itinerant character of the 4$f$ electrons in CeRu$_2$
\cite{SekiyamaA:Probsc}.

In the following, we will move on to the review of recent ARPES
results from several materials, such as Sr$_2$RuO$_4$,
2$H$-NbSe$_2$, Be(0001), and Mo(110). These examples will be used
to illustrate the capability of this technique and some of the
specific issues that one can investigate in detail by ARPES. In
particular, these test cases will demonstrate that, by taking full
advantage of the momentum and energy resolution as well as of the
photon energy range nowadays available, state-of-the-art ARPES is
a unique tool for {\it momentum space microscopy}.

\subsection{Sr$_2$RuO$_4$: Bands and Fermi Surface}

To illustrate how one can study electronic bands and Fermi
surfaces by ARPES, and how critical the improvement in resolution
has been in this regard, the novel superconductor Sr$_2$RuO$_4$ is
a particularly good example. Its low-energy electronic structure,
as predicted by band-structure calculations is characterized by
three bands crossing the chemical potential
\cite{SinghDJ:RelS24,OguchiT:Elebst}. These define a complex Fermi
surface comprised of two electron pockets and one hole pocket
(Fig.\,\ref{srofs}d), which have been clearly observed in de
Haas-van Alphen experiments
\cite{MackenzieAP:Quaotl,BergemannC:DetttF}. On the other hand,
early photoemission measurements suggested a different topology
\cite{YokoyaT:Ang-re,YokayaT:ExtVHs,LuDH:Fersev}, which generated
a certain degree of controversy in the field
\cite{PuchkovAV:ARPrS2}. This issue was conclusively resolved only
by taking advantage of the high energy and momentum resolution of
the `new generation' of ARPES data: it was then recognized that a
surface reconstruction \cite{MatzdorfR:Fersbl} and, in turn, the
detection of several direct and folded surface bands were
responsible for the conflicting interpretations
\cite{DamascelliA:Fersss,DamascelliA:FersS2,DamascelliA:1,shenkm:1}.
Fig.\,\ref{srofs}a,b show high resolution ARPES data ($\Delta
E\!=\!14$ meV, $\Delta k\!=\!1.5$\% of the zone edge) taken at
10\,K with 28\,eV photons on a Sr$_2$RuO$_4$ single crystal
cleaved at 180 K (for Sr$_2$RuO$_4$, as recently discovered,
high-temperature cleaving suppresses the reconstructed-surface
contributions to the photoemission signal and allows one to
isolate the bulk electronic structure \cite{DamascelliA:Fersss}).
Many well defined quasiparticle peaks disperse towards the Fermi
energy and disappear upon crossing $E_F$. A Fermi energy intensity
map (Fig.\,\ref{srofs}c) can then be obtained by integrating the
spectra over a narrow energy window about $E_F$ ($\pm10$ meV). As
the spectral function (multiplied by the Fermi function) reaches
its maximum at $E_F$ when a band crosses the Fermi energy, the
Fermi surface is identified by the local maxima of the intensity
map. Following this method, the three sheets of Fermi surface are
clearly resolved and are in excellent agreement with the
theoretical calculations (Fig.\,\ref{srofs}d).

\subsection{2$H$-NbSe$_2$: Superconducting Gap}

\begin{figure}[t!]
\centerline{\epsfig{figure=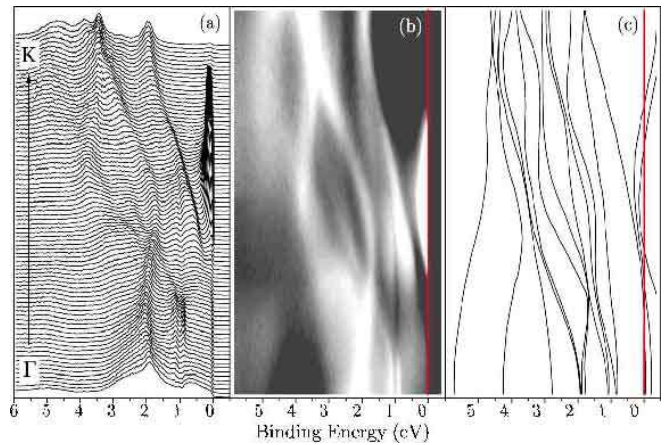,width=1\linewidth,clip=}}
\caption{(Color). (a) 2$H$-NbSe$_2$ ARPES spectra (measured at
20\,K with 21.2\,eV photons), (b) corresponding image plot, and
(c) band structure calculations along $\Gamma$-$K$ (from
Ref.\,\onlinecite{Damascelli:nbse}).}\label{nbseandrea}\end{figure}

2$H$-NbSe$_2$ is an interesting quasi two-dimensional system
exhibiting a charge-density wave phase transition at approximately
33\,K, and a phonon-mediated superconducting phase transition at
7.2\,K. As indicated by band structure calculations
\cite{corcoran:1}, the valence-band electronic structure is
characterized by a manifold of dispersive bands in a 6\,eV range
below the Fermi energy. At low energy, three dispersive bands are
expected to cross the chemical potential and define three sheets
of Fermi surface in the hexagonal Brillouin zone. Both the band
manyfold and the Fermi surface topology have been studied in great
detail by ARPES; exception made for a weak energy renormalization,
the normal-state experimental data are in extremely good agreement
with the results of theoretical calculations (as shown in
Fig.\,\ref{nbseandrea}, where ARPES spectra and band structure
calculations are compared for the $\Gamma$-$K$ high symmetry
direction). As for the low temperature charge-density wave phase,
despite the intense effort no agrement has been reached yet on the
driving force responsible for the transition
\cite{pillo:1,StraubT:nbse}.

Owing to the great improvement in energy and momentum resolution,
it has now become possible to study by ARPES also the momentum and
temperature dependence of the superconducting gap on low-$T_c$
materials (until recently, experiments of this kind could been
performed only for the much larger d-wave gap of the high-$T_c$
superconductors \cite{Damascelli:RMP}).
\begin{figure}[b!]
\centerline{\epsfig{figure=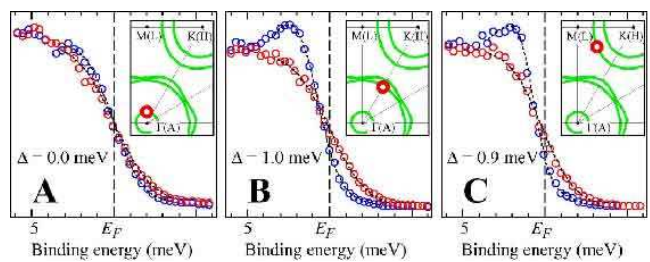,width=1\linewidth,clip=}}
\caption{(Color). Normal (10\,K, red) and superconducting state
(5.3\,K, blue) ARPES spectra from 2$H$-NbSe$_2$, measured at
$k$-points belonging to the three different sheets of Fermi
surface (see insets). The value of the superconducting gap used to
fit the data is indicated in each panel (from
Ref.\,\onlinecite{YokoyaT:nbse}).} \label{yokoya}\end{figure}
The data presented in Fig.\,\ref{yokoya}, which are one of the
most impressive examples of combined high energy and momentum
resolution in ARPES experiments on solid samples (i.e., $\Delta
E\!=\!2.5$\,meV and $\Delta k\!=\!0.2^\circ$), provide direct
evidence for Fermi surface sheet-dependent superconductivity in
2$H$-NbSe$_2$ \cite{YokoyaT:nbse}. A superconducting gap of about
1\,meV was successfully detected along two of the normal-state
Fermi surface sheets, but not along the third one. In fact, the
opening of the gap is directly evidenced in Fig.\,\ref{yokoya}b
and\,\ref{yokoya}c by the shift to high binding energies of the
5.3\,K spectra leading-edge midpoint (which is instead located at
$E_F$ at 10\,K, as expected for a metal), and by the simultaneous
appearance of a peak below $E_F$ (which reflects the piling up of
the density of states due to the gap opening). This behavior is
absent for the inner Fermi surface pocket (Fig.\,\ref{yokoya}a).

\subsection{Self Energy and Collective Modes}

As discussed in Sec.\,\ref{sec:spectral}, the introduction of the
electron self energy $\Sigma({\bf
k},\omega)\!=\!\Sigma^{\prime}({\bf
k},\omega)\!+\!i\Sigma^{\prime\prime}({\bf k},\omega)$ is a
powerful way to account for many-body correlations in solids. Its
real and imaginary parts correspond, respectively, to the energy
renormalization with respect to the bare band energy
$\epsilon_{\bf k}$ and to the finite lifetime of the
quasiparticles in the interacting system. Owing to the energy and
momentum resolution nowadays achievable, both components of the
self energy can be in principle estimated very accurately from the
analysis of the ARPES intensity in terms of {\it energy
distribution curves} (EDCs) and/or {\it momentum distribution
curves} (MDCs), which is one of the aspects that make ARPES such a
powerful tool for the investigation of complex materials. In some
cases the MDC analysis may be more effective than the analysis of
the EDCs in extracting information on the self energy. In fact,
EDCs are typically characterized by a complex lineshape
(Fig.\,\ref{valla1}) because of the nontrivial $\omega$ dependence
of the self energy, the presence of additional background, and the
low-energy cutoff due to the Fermi function. Furthermore, as
evidenced by the generic expression for the spectral function
$A({\bf k},\omega)$ in Eq.\,\ref{eq:akw1}, the EDC peak position
is determined by $\Sigma^{\prime}({\bf k},\omega)$ as well as
$\Sigma^{\prime\prime}({\bf k},\omega)$, because both terms are
strongly energy dependent. On the other hand, if the self energy
is independent of $k$ normal to the Fermi surface (and the matrix
elements are a slowly-varying function of $k$), then the
corresponding MDCs are Lorentzians centered at
$k\!=\!k_F\!+\![\omega\!-\!\Sigma^\prime(\omega)]/v_F^0$ with FWHM
given by $2\Sigma^{\prime\prime}(\omega)/v_F^0$, where $v_F^0$ is
the bare Fermi velocity normal to the Fermi surface [this is
obtained by approximating $\epsilon_{\bf
k}\!\simeq\!v_F^0(k\!-\!k_F)$ in Eq.\,\ref{eq:akw1}]. Lorentzian
lineshapes were indeed observed for the MDCs (Fig.\,\ref{valla1}).

\begin{figure}[b!]
\centerline{\epsfig{figure=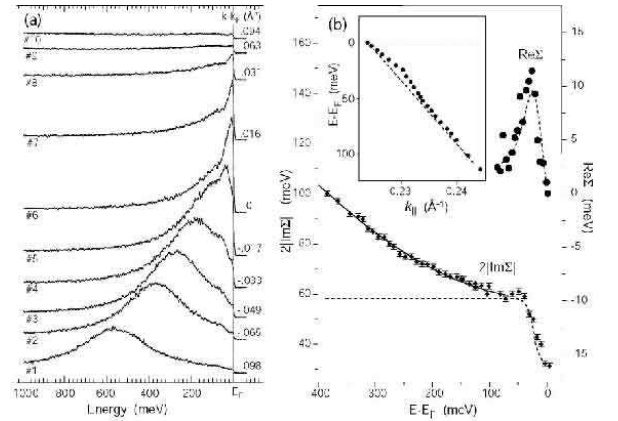,clip=,angle=0,width=0.9\linewidth}}
\caption{(a) ARPES spectra for the Be(0001) surface state (from
Ref.\,\onlinecite{HengsbergerM:Phossc}). (b) Self energy estimated
from the Mo(110) surface state ARPES spectra, and corresponding
quasiparticle dispersion (inset). Calculated electron-phonon
contributions to the real and imaginary part of $\Sigma({\bf
k},\omega)$ are indicated by dotted and dashed lines, respectively
(the latter was offset by 26\,meV to account for impurity
scattering). From Ref. \onlinecite{VallaT:Man-bo}.} \label{baer}
\end{figure}

As an example of this kind of analysis we will briefly discuss the
case of electron-phonon coupling on metallic surfaces, for which
the established theoretical formalism can be applied very
effectively
\cite{BalasubramanianT:Larvte,HofmannP:Ele-la,HengsbergerM:Phossc,HengsbergerM:Ele-ph,VallaT:Eviqcb,LaShellS:Nonstp,RotenbergE:Coubav}.
The electron-phonon interaction involving surface phonons and the
$\overline{\Gamma}$-surface state on the Be(0001) surface was
investigated by two groups, and qualitatively similar conclusion
were drawn
\cite{BalasubramanianT:Larvte,HengsbergerM:Phossc,HengsbergerM:Ele-ph,LaShellS:Nonstp}.
Fig.\,\ref{baer}a shows results for the Be(0001) surface state
along the $\overline{\Gamma M}$ direction of the surface Brillouin
zone; a feature is seen dispersing towards the Fermi level
\cite{HengsbergerM:Phossc} . Close to $E_F$ the spectral function
exhibits a complex structure characterized by a broad hump and a
sharp peak, with the latter being confined to within an energy
range given by the typical bandwidth $\omega_{ph}$ of the surface
phonons. This behavior corresponds to a `two-branch' splitting of
the near-$E_F$ dispersion, with a transfer of spectral weight
between the two branches as a function of binding energy. While
the high-energy dispersion is representative of the bare
quasiparticles, at low energy the dispersion is renormalized by
the electron-phonon interaction (this behavior is shown, for a
similar electron-phonon coupled system, in the inset of
Fig.\,\ref{baer}b).
In other words, the weaker dispersion observed at energies smaller
than $\omega_{ph}$ describes dressed quasiparticles with an
effective mass enhanced by a factor of $(1\!+\!\lambda)$, where
$\lambda$ is the electron-phonon coupling parameter
\cite{Ashcroft:1}. The latter can also be estimated from the ratio
of renormalized (${\bf v}_{{\bf k}}^{ }$) and bare (${\bf v}_{{\bf
k}}^{0}$) quasiparticle velocities, according to the relation
${\bf v}_{\bf k}^{ }\!=\!\hbar^{-1}\partial\varepsilon_{\bf
k}/\partial {\bf k} \!=\!(1\!+\!\lambda)^{-1}{\bf v}_{\bf k}^{0}$.
This way, for the data presented in Fig.\,\ref{baer}a the value
$\lambda\!=\!1.18$ was obtained (alternatively $\lambda$ can also
be estimated from the temperature dependence of the linewidth near
$E_F$ \cite{BalasubramanianT:Larvte}).

A similar example of electron-phonon coupled system is the surface
of Mo(110) \cite{VallaT:Man-bo}. In this case, the real and
imaginary part of the self energy shown in Fig.\,\ref{baer}b were
obtained directly from the EDC analysis: $\Sigma^{\prime\prime}$
corresponds to the EDC width and $\Sigma^{\prime}$ to the
difference between the observed quasiparticle dispersion and a
straight line approximating the dispersion of the non-interacting
system (Fig.\,\ref{baer}b, inset). The step-like change at 30 meV
in $\Sigma^{\prime\prime}$ is interpreted as the phonon
contribution (dashed line) and the parabolic part at higher
energies is attributed to electron-electron interactions. The
phonon contribution to the real part of the self energy is
calculated from the Kramers-Kronig relations (see
Sec.\,\ref{sec:spectral}) and agrees well with the data (dotted
line). As an additional confirmation of the electron-phonon
description, it was noted that the temperature dependence of the
scattering rate is well reproduced by the calculations
\cite{VallaT:Man-bo}.

\bibliographystyle{apsrev}
\bibliography{ARPES_Intro}


\end{document}